\documentclass[%
 preprint,
 titlepage,
 amsmath,amssymb,
 aps,
nofootinbib
 ]{revtex4-2}

\usepackage{graphicx}
\usepackage{dcolumn}
\usepackage{bm}
\usepackage{hyperref}
\hypersetup{colorlinks=true, citecolor=blue, urlcolor=blue, linkcolor=blue}

\begin{document}


\title{Inflation models selected by the swampland distance conjecture  with the Lyth bound}

\author{Yuma S. Furuta$^{1,2}$, Yuta Hamada$^{1,2}$ and Kazunori Kohri$^{3,4,2,5}$}

\affiliation{
$^1$ School of High Energy Accelerator Science, Graduate University for Advanced Studies (SOKENDAI), 1-1 Oho, Tsukuba, Ibaraki 305-0801, Japan}
\affiliation{
$^2$ Theory Center, IPNS, High Energy Accelerator Research Organization (KEK), 1-1 Oho, Tsukuba, Ibaraki 305-0801, Japan}
\affiliation{
$^3$ Division of Science, National Astronomical Observatory of Japan, 2-21-1 Osawa, Mitaka, Tokyo 181-8588, Japan}
\affiliation{
$^4$ School of Physical Sciences, Graduate University for Advanced Studies (SOKENDAI), 2-21-1 Osawa, Mitaka, Tokyo 181-8588, Japan}
\affiliation{
$^5$ Kavli IPMU (WPI), UTIAS, The University of Tokyo, Kashiwa, Chiba 277-8583, Japan}

\date{\today}

\begin{abstract}
We investigate the extent to which the swampland conjecture can be
employed to constrain large-field inflationary models from the
perspective of quantum gravity consistency. In particular, we focus
on the swampland distance conjecture, which imposes an upper bound
on the amplitude of primordial gravitational waves predicted by
large-field inflation scenarios. This provides a striking contrast
with the well-known Lyth bound, which yields a lower bound on the
tensor-to-scalar ratio in such models. The two bounds thus play
complementary roles in assessing the viability of inflationary
scenarios. We demonstrate that, for certain representative
large-field inflation models, the swampland distance conjecture 
alone can impose more stringent upper limits on the tensor-to-scalar
ratio than current observational constraints from the cosmic
microwave background. These findings highlight the utility of
swampland criteria as a theoretical discriminator among competing
inflationary models, independent of empirical data.
\end{abstract}

\vspace*{-5em}
\maketitle

\section{Introduction}
\label{sec:intro}

Cosmic inflation offers elegant resolutions to several fundamental
problems in standard cosmology, including the horizon problem, the
flatness problem, the monopole problem, and the origin of primordial
density perturbations~\cite{Ellis:2023wic}. As such, uncovering the
underlying mechanism responsible for inflation remains one of the most
pressing and central challenges in modern theoretical
cosmology. Despite its theoretical success, the precise nature of the
scalar field $\phi$, commonly referred to as the inflaton field that
is believed to have driven inflationary dynamics has yet to be
established. Neither the energy scale at which inflation occurred nor
its precise epoch in cosmic history has been directly probed by
current observational data. Observations of the cosmic microwave
background (CMB), particularly its temperature anisotropies and
polarization patterns, have so far yielded only upper bounds on the
inflationary energy scale, constraining it to be below approximately
$10^{16}$ GeV. This constraint is typically expressed in terms of the
tensor-to-scalar ratio $r$, with the most stringent current bound
being $r < 0.036$~\cite{BICEP:2021xfz, Tristram:2021tvh}.

On the theoretical front, there exist complementary efforts to
constrain the energy scale of inflation based on internal consistency
relations derived from purely theoretic arguments. One such relation
is the Lyth bound, which applies to slowroll models with
super-Planckian field excursions, and provides a model-independent
lower limit on $r$. However, even when considering both observational
upper bounds and theoretical lower bounds, significant uncertainties
remain regarding the model-dependent functional form of the inflaton
potential $V(\phi)$ and the total field excursion $\Delta \phi$ during
inflation.

Furthermore, recent developments in quantum gravity, particularly in the context of string theory, have motivated theoretical constraints on effective field theories via the so-called swampland conjectures~\cite{Vafa:2005ui}. In particular, the swampland distance conjecture  (SDC)~\cite{Ooguri:2006in} posits that effective field theories with trans-Planckian field excursions are incompatible with a consistent UV completion in quantum gravity.\footnote{See also Ref.~\cite{Cribiori:2025oek} for bounds on the field excursion from the UV/IR mixing motivated by the holography and entropy bounds.} The SDC, if valid, would place nontrivial theoretical constraints on inflationary model building, potentially ruling out large classes of models otherwise consistent with observations.

In this work, we present a detailed calculation of the
tensor-to-scalar ratio as a function of $\Delta \phi$ for several
representative large-field inflation models. Under specific
assumptions, we demonstrate that the swampland distance conjecture 
can, in certain cases, impose more stringent bounds than those derived
from current observational data.

The structure of the paper is as follows: In Sec. II, we review the
swampland conjecture and its implications for inflationary
dynamics. Section III outlines the Lyth bound. In Sec. IV, we
introduce a selection of concrete large-field inflation
models. Section V presents detailed model-dependent predictions for
the tensor-to-scalar ratio and discusses the interplay between
observational constraints and theoretical bounds. We summarize our
findings and conclude in Sec. VI. Throughout this paper, we adopt
natural units where $\hbar = c = 1$.

\section{Bound from the swampland distance conjecture }
\label{sec:swamp}

\begin{figure}[!h]
  \centering
  \includegraphics[width=0.8\columnwidth]{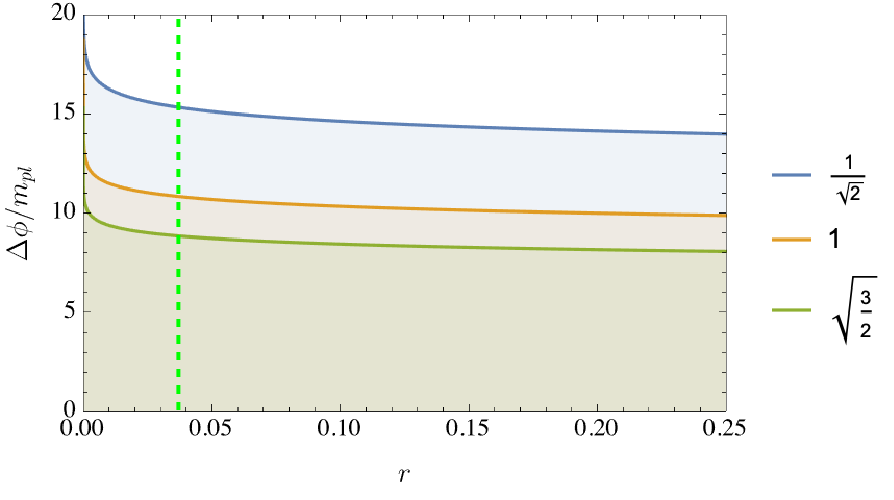}
  \caption{Upper bounds on the tensor-to-scalar ratio $r$ as a
    function of the excursion distance of the inflaton
    $\Delta\phi$. Regarding the reference values for
    $\lambda_{\mathrm{dc}}$, we take $1$, $\sqrt{\frac{3}{2}}$, and
    the minimum $\frac{1}{\sqrt{2}}$ in the d=4 dimensions. The
    vertical green dashed line is the observational upper bound on the
    tensor-to-scalar ratio, $r < 0.036$ (95$\%$ CL) by the
    observational data of the cosmic microwave background
    (CMB)~\cite{BICEP:2021xfz,Tristram:2021tvh}.}
  \label{fig:sdc}
\end{figure}

By using SDC, the authors of Refs.~\cite{Hebecker:2018vxz,Scalisi:2018eaz} argue that
the Hubble expansion rate 
during the primordial inflation should be smaller than the
cutoff scale,
$H \leq m_{\rm pl} e^{- \lambda_{\mathrm{dc}} \Delta\phi/ m_{\rm
    pl}}$.
Here $\lambda_{\mathrm{dc}}$ is the exponential rate at which an infinite tower of states becomes light,
$m_{\rm pl}$ is the (reduced) Planck mass ($\simeq2.4 \times 10^{18}$~GeV), and $\Delta\phi$ is the excursion distance of the inflaton $\phi$ during the inflation.    
The constraint can be viewed as the upper bound on the field excursion
\begin{equation}
\label{eq:sdc}
\frac{\Delta\phi}{m_{\rm pl}} \leq \frac{1}{\lambda_{\mathrm{dc}}} \log \left(\frac{m_{\rm pl}}{H}\right).
\end{equation}
When we rewrite this
inequality as a relation between $\Delta\phi$ and the tensor-to-scalar
ratio $r$, we obtain
\begin{equation}
\label{eq:sdcr}
\frac{\Delta\phi}{m_{\rm pl}} \leq \frac{1}{2\lambda_{\mathrm{dc}}} \log \left(\frac{2}{\pi^2 A_s r}\right),
\end{equation}
where $A_s$ is the amplitude of the curvature perturbation
($\sim 2.1 \times 10^{-9}$) produced by the inflation at the horizon crossing of a large scale (e.g.,
$\sim 0.05\ {\rm Mpc}^{-1}$)~\cite{Planck:2018vyg}. Based on the emergent string conjecture~\cite{Lee:2019wij}, the sharpened distance conjecture~\cite{Etheredge:2022opl, Heidenreich:2019zkl} (see also~\cite{Agmon:2022thq}) puts a lower bound on the parameter
$\lambda_{\mathrm{dc}}$ to be
$\lambda_{\mathrm{dc}} \geq\frac{1}{\sqrt{d-2}}$, where $d$ is the number of dimensions. For concrete values of
$\lambda_{\mathrm{dc}}$, the following three are chosen in this paper.
\begin{equation}
\lambda_{\mathrm{dc}}=\left\{\begin{array}{cl}
1 & \text { (a reference value), } \\
\sqrt{\frac{D-2}{(D-d)(d-2)}}
=\sqrt{\frac{3}{2}} & \text { [Kaluza Klein (KK) tower], } \\
\frac{1}{\sqrt{d-2}}=\frac{1}{\sqrt{2}} & \text { (string tower). }
\end{array}\right.
\end{equation}
Here $D$ is the dimension before the compactification for the KK tower, and we take $D=d+1$ and $d=4$. 
The first one,
$\lambda_{\mathrm{dc}}=1$, is just for reference
for simplicity. The second one
is derived naturally from the dimension reduction, and it is the same as the case for the KK tower. The third one
is the minimum value of
$\lambda_{\mathrm{dc}}$ and corresponds to the case where the string tower is the lightest~\cite{Etheredge:2022opl, Heidenreich:2019zkl,Agmon:2022thq}. The upper bounds on $\Delta \phi$ as a function of $r$ are illustrated in
Fig.~\ref{fig:sdc} for three cases.
We note that, in a controlled scenario of string inflation (see, e.g., Refs.~\cite{Baumann:2014nda,Cicoli:2023opf} for reviews), the KK tower is always lighter than the string tower. Therefore, it is natural to take the case of the KK tower, which puts the stringent bound on $\Delta \phi$.

The SDC up to this point provides the upper
bounds on the tensor-to-scalar ratio $r$ as a function of the
excursion distance of the field $\Delta\phi$. In the next section, we
will look at the Lyth bound, which provides the lower bound on the
tensor-to-scalar ratio $r$ as a function of $\Delta\phi$.

\section{Lyth bound}
\label{sec:LB}

In this section, we consider possible constraints imposed on the model
parameters of the inflation by the Lyth bound. We discuss typical
large-field models of the slowroll inflation in which the inflation is
induced by the inflaton field slowly rolling on a flat potential in
the beginning. In this case, the inflation ends by the breakdown of
the conditions for the slowroll at a late time due to the fast rolling
on the potential which becomes steeper than the one in the
beginning. Then, the Lyth bound gives an upper bound on the
tensor-to-scalar ratio $r$ as a function of the field excursion
$\Delta\phi$.  Here, we define the first and second slowroll
parameters as a function of $\phi$ as follows.
\begin{eqnarray}
\label{eq:slowrollparam}
\epsilon (\phi) \equiv \frac{m^2_{\rm pl}}{2} \left ( \frac{V,_{\phi}}{V} \right )^2,\\
\eta (\phi) \equiv m^2_{\rm pl} \left ( \frac{V,_{\phi\phi}}{V} \right )^2.
\end{eqnarray}
where the subscript $,{\phi}$ denotes the differentiation with
respect to $\phi$. The $e$-folding number during the inflation is defined by
\begin{eqnarray}
\label{eq:efoldingnum}
N(t) &\equiv& \ln \frac{a_{\rm end}}{a(t)} = \int ^{a_{\rm end}} _{a(t)} \frac{da}{a} = \int ^{t_{\rm end}} _t H dt,
%
\end{eqnarray}
where the subscript \textit{end} represents the value at the end of the
inflation $t = t_{\rm end}$.  

To solve both the horizon problem and the flatness problem, the
$e$-folding number $N(t_{\rm CMB})$ between a kind of the initial time
$t=t_{\rm CMB}$,~\footnote{Correctly, at $t=t_{\rm CMB}$, the mode,
  which produced the fluctuation of the CMB, exited the horizon. This
  is not the initial time.  The actual initial time $t_{\rm initial}$
  should be shorter than $t_{\rm initial} \le t_{\rm CMB}$} 
and the end of inflation ($t=t_{\rm end}$) must be greater than 47--62. The
upper limit ($N$=62) comes from the upper bound on the energy scale of
the potential of the inflation $V^{1/4} \lesssim 10^{16}$~GeV, which
corresponds to the upper bound on the tensor-to-scalar ratio
$r<0.036$~\cite{BICEP:2021xfz,Tristram:2021tvh}. On the other hand,
the lower limit ($N$ = 47) comes from the most conservative lower
bound on the reheating temperature after inflation
($T_R > 4$~MeV~\cite{Hasegawa:2019jsa}). This is obtained from the
conditions for the successful big-bang nucleosynthesis in terms of the
thermalization of the background neutrino and keeping the neutron to
proton ratio ($n/p$) unchanged even by the scatterings of the emitted
high-energy particles off the background
particles~\cite{Kawasaki:1999na,Kawasaki:2000en,Hannestad:2004px,Ichikawa:2005vw,deSalas:2015glj,Hasegawa:2019jsa}.

\begin{figure}[!h]
  \centering
  \includegraphics[width=0.8\columnwidth]{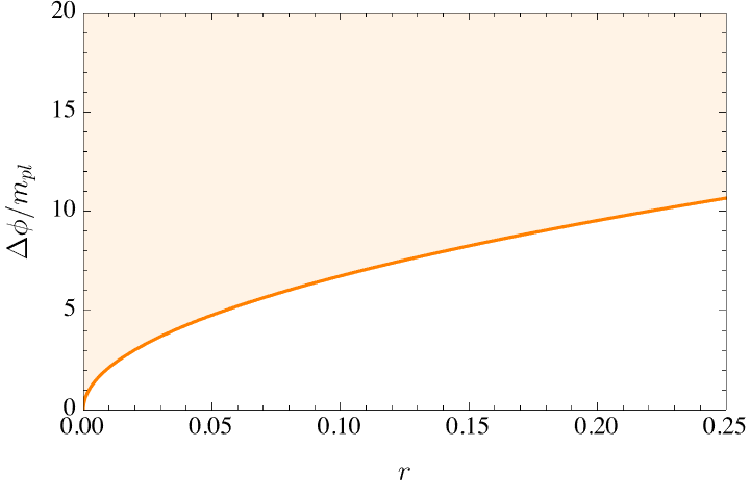}
  \caption{Lower bound on $\Delta \phi$ as a function of $r$ from the Lyth bound. Here we put $\Delta \phi = 60$ for a representative value in large field models.}
  \label{fig:lb}
\end{figure}

Furthermore, expressing this relation as a function of the scalar
field $\phi$, we have
\begin{eqnarray}
\label{eq:efoldingnumSc}
N(\phi_{\rm CMB}) = \int ^{\phi_{\rm CMB}} _{\phi_{\rm end}} \frac{V}{V,_{\phi}} \frac{d\phi}{M^2_p}.
\end{eqnarray}
By the definition, the tensor-to-scalar ratio $r$ is also expressed by
\begin{eqnarray}
\label{eq:r}
r \equiv \frac{P_T}{P_\zeta} = 16\epsilon,
\end{eqnarray}
where $P_T = \frac{2 V}{ 3 \pi^2 m_{\rm pl}^4}$ is the tensor perturbation, and
$P_\zeta = \frac{V}{24 \pi^2 m_{\rm pl}^4 \epsilon}$ is the scalar curvature
perturbation. Regarding the Lyth bound, expressing the $e$-foldings
number (\ref{eq:efoldingnumSc}) using the slowroll parameter
(\ref{eq:slowrollparam}), we obtain
\begin{eqnarray}
\label{eq:efoldsr}
N(\phi_{\rm CMB}) = \int ^{\phi_{\rm CMB}} _{\phi_{\rm end}} 
\frac{1}{ \sqrt{2\epsilon(\phi)}} \frac{d\phi}{m_{\rm pl}}.
\end{eqnarray}
Inside the integrand, if we took an initial value as the representative
of each variable, e.g., $\epsilon(\phi) = \epsilon(\phi_{\rm CMB})$ at
the beginning $t = t_{\rm CMB}$, because of the
inequality $\epsilon(\phi_{\rm end}) > \epsilon(\phi_{\rm CMB})$, we obtain
\begin{eqnarray}
\label{eq:lb}
r \leq 2.2\times10^{-3}\left(\frac{\Delta N}{60}\right)^{-2}\left(\frac{\Delta \phi}{m_{\rm pl}}\right)^2,
\end{eqnarray}
with  $\Delta N = N_{\rm end} - N_{\rm CMB}$ and
$\Delta \phi = |\phi_{\rm CMB} - \phi_{\rm end}|$, respectively. This inequality is called the Lyth bound
In Fig.~\ref{fig:lb}, we plot the Lyth bound which gives the lower bound on $\Delta \phi$ as a function of $r$.

\section{Inflation Models}
\label{sec:inflation}

In this section, we discuss the four typical large-field inflation models as follows.

\subsection{Chaotic inflation}
\label{ssec:chaotic}

First, we consider the models of chaotic inflation~\cite{LINDE1983177}
which has the simplest potential forms and allows the scalar field to
have a very large initial value more than the Planck mass. The
potential is represented by the monomial function of $\phi$,
\begin{equation}
\label{eq:chao}
V_{\rm{C}}(\phi) = V_0 \left(\frac{\phi}{m_{\rm pl}}\right)^p,
\end{equation}
where $p$ is an exponent. Then, the tensor-to-scalar ratio $r$ is
analytically derived to be
\begin{eqnarray}
\label{eq:rcha}
r =  8\left(\frac{\phi}{p m_{\rm pl}}\right)^{-2},
\end{eqnarray}
where $\phi$ depends on $p$ and $N$, i.e., $\phi = \phi(p,N)$.

\subsection{Natural inflation}
\label{ssec:natural}

Next, we discuss the natural inflation model~\cite{Adams:1992bn},
which utilizes a pseudo-Nambu-Goldstone boson as the inflaton. The
potential of the models of natural inflation is given by
\begin{equation}
\label{eq:nat}
V_{\mbox{N}}(\phi) = V_0 \left(1-\cos{\left(\frac{\phi}{F}\right)}\right),
\end{equation}
where $F$ is a parameter (a decay constant) parametrizing the
periodicity, and the gentle slope of the potential, which is required for
inflation to be naturally realized. Then, the
tensor-to-scalar ratio $r$ is expressed by
\begin{eqnarray}
\label{eq:rnat}
r =  8\left(\frac{\sin(\phi/F)}{1-\cos(\phi/F)}\right)^{2}\left(\frac{F}{m_{\rm pl}}\right)^{-2},
\end{eqnarray}
where $\phi$ depends on $F$ and $N$, i.e., $\phi = \phi(F,N)$.

\subsection{Hilltop inflation}
\label{ssec:hilltop}

Concrete forms of the models of hilltop inflation were studied by
e.g.,
Refs.~\cite{Boubekeur:2005zm,Kohri:2007gq,Kohri:2014rja,Lin:2018rnx,Lin:2019fdk,Dimopoulos:2020kol},
which describes inflation starting near the top of the hill of the
potential. It has been shown that hilltop potentials can be easily
obtained from the F term or the D term in
supersymmetry~\cite{Boubekeur:2005zm,Kohri:2007gq}.

The quadratic hilltop inflation is parametrized by
\begin{equation}\label{eq:natural}
V_{\mbox{H}}(\phi) = V_0 (1-\phi^4/\mu^4_4),
\end{equation}
where $\mu_4$ is a parameter. Then, the tensor-to-scalar ratio $r$ is
represented by
\begin{eqnarray}\label{eq:rhill}
r =  8\left(\frac{4 m_{\rm pl} \phi^3}{\mu_4^4-\phi^4}\right)^2,
\end{eqnarray}
where $\phi$ depends on $\mu_4$ and $N$, i.e., $\phi = \phi(\mu_4,N)$.

\subsection{$\alpha$ attractors}
\label{ssec:alp}

Finally, we discuss the models of $\alpha$ attractors~\cite{Ferrara:2013kca,Kallosh:2013maa}. These models have been
proposed and studied in the context of supergravity.  For example, the potential for the $T$ model and $E$ model of the $\alpha$ attractors is given by
\begin{eqnarray}
\label{eq:attractor}
V_{\mbox{$\alpha$}}(\phi) = V_0 \tanh^q\left(\frac{\phi}{\sqrt{6\alpha}m_{\rm pl}}\right),\\
V_{\mbox{$\alpha$}}(\phi) = V_0 \left( 1- e^{-\sqrt{2}\phi / \sqrt{3\alpha}m_{\rm pl}}  \right)^q,
\end{eqnarray}
where $\alpha$ is the parameter. The $\alpha$ of two models takes the same range as mentioned in the next section, so we use the same representation, $\alpha$.
In this paper, we adopt $q=2$ for a representative value. By transforming to a canonically
normalized field into the Einstein frame, an exponentially flat region
is obtained. Then, the tensor-to-scalar ratio of the $T$ model and $E$ model is expressed by
\begin{eqnarray}\label{eq:rcha}
r =  \frac{64}{3\alpha}\left(\sinh\frac{\phi}{\sqrt{6\alpha m_{\rm pl}}}\cosh\frac{\phi}{\sqrt{6\alpha m_{\rm pl}}}\right)^{-2},\\
r =  \frac{64}{3\alpha} \frac{e^{-2\sqrt{2}\phi / \sqrt{3\alpha}m_{\rm pl}} }{\left( 1- e^{-\sqrt{2}\phi / \sqrt{3\alpha}m_{\rm pl}}  \right)^2},
\end{eqnarray}
where $\phi$ depends on $\alpha$ and $N$, i.e., $\phi = \phi(\alpha,N)$.

\section{Combined limits from swampland distance conjecture  and Lyth bound}
\label{sec:swdl}

In this section, we consider the region enclosed by the swampland distance
conjecture in (\ref{eq:sdcr}) and the Lyth bound in (\ref{eq:lb}) in the
2D plane of the field distance $\Delta\phi$ and the tensor-to-scalar
ratio $r$. 
Among the representative four inflation models described in
Sec.~\ref{sec:inflation}, we showed the relationship between the
field distance and the tensor-to-scalar ratio which are related to each
other through the $e$-foldings number $N$ given in
Eq.~(\ref{eq:efoldsr}). We plot the value of $r$ as a function of
$\Delta \phi$ predicted in each model on the same plane and compare it
with the region enclosed by the swampland distance conjecture  and the
Lyth bound. Here, the parameters adopted in this
analysis are shown in Table~\ref{tab:table1}.
\begin{table}[b]
\caption{\label{tab:table1}%
Parameter ranges for each inflation model.
}
\begin{ruledtabular}
\begin{tabular}{lcdr}
\textrm{Inflation model}&
\textrm{Parameter range}
\\
\colrule
Chaotic &    $0 < p < 4$\\
Natural &    $0.3 < \log_{10}\left(F/m_{\rm pl}\right) < 2.5$\\
Hilltop &    $-2 < \log_{10}\left(\mu_4/m_{\rm pl}\right) < 2$\\
$\alpha$  attractor &    $-2 < \log_{10}\left(\alpha\right) < 4$,\\
\end{tabular}
\end{ruledtabular}
\end{table}
which refers to the ranges of the parameters studied in the paper of
the Planck collaboration in 2018~\cite{Planck:2018jri}.

\begin{figure}[!h]
  \centering
  \includegraphics[width=0.8\columnwidth]{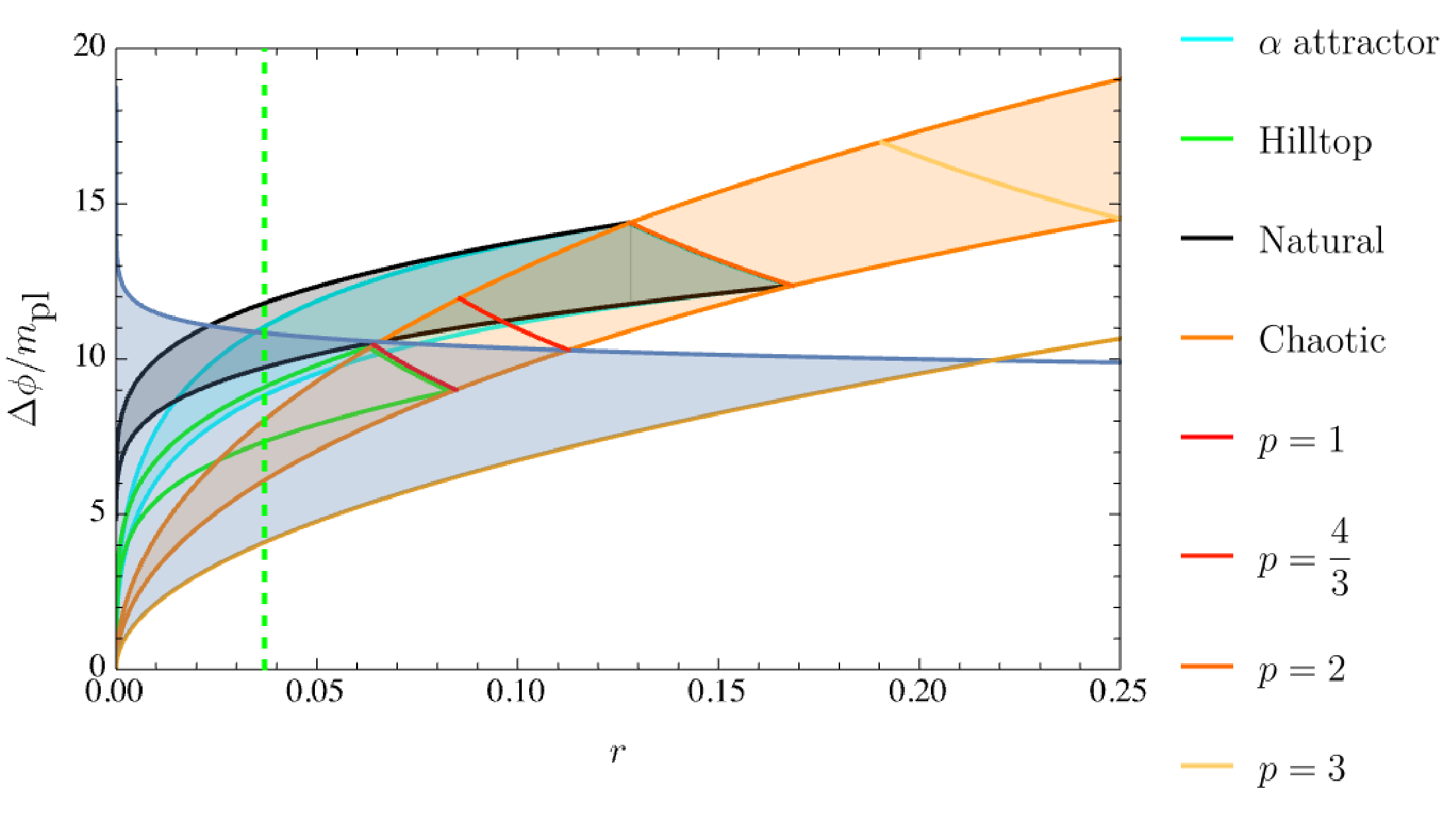}
  \caption{Allowed regions for four inflation models enclosed by the
    swampland distance conjecture  (blue) and the Lyth bound (brown). We adopted the
    parameter $\lambda_{\mathrm{dc}} = 1$ for the swampland Distance
    Conjecture. For the theoretical calculations, we plotted the cases
    for chaotic inflation with the red lines for $p=1, 4/3, 2, 3$,
    natural inflation (black), hilltop inflation (green) and
    $\alpha$ attractors (cyan). Two lines mean $N$=47 (lower one)
    and $N$=62 (upper one), respectively. The vertical green dashed line is the
    observational upper bound on $r$ by the data of the
    cosmic microwave background
    (CMB)~\cite{BICEP:2021xfz,Tristram:2021tvh}.}
  \label{fig:sdc_lb_e1}
\end{figure}

The results are shown in Fig.~\ref{fig:sdc_lb_e1}. The parameter
$\lambda_{\mathrm{dc}}$ of the swampland distance conjecture is taken
to be 1. For some values within the parameter range of each inflation
model, the number of $e$-foldings is taken from 47 to 62. For example,
$p=1$ for the chaotic inflation model is the darkest red line plotted
traversing the orange chaotic region in
Fig.~\ref{fig:sdc_lb_e1}. This red line is obtained by fixing the
parameter $p=1$ and plotting the $e$-foldings number from 47 to
62. The region obtained when the parameter $p$ is varied within the
range of Table~\ref{tab:table1} is the region enclosed by the light
orange color. The two longest orange curves are obtained by fixing the
$e$-foldings number to be 62 (upper one) or 47 (lower one),
respectively, and varying the parameter $p$. Similarly, the plot for
natural inflation is the black region enclosed by the black
curves. The plot for hilltop inflation is the green region enclosed by
the green curves. The plot for the $\alpha$ attractor is the light
blue region enclosed by the light blue curves. 
The regions of the $T$ model and $E$ model are plotted on an extremely close area, so we combined these as the figure of the $\alpha$ attractor.
As is clear from the
figure, the natural inflation model and the $\alpha$ attractor
inflation model coincide with the case of $p=2$ for the chaotic
inflation model in the large parameter regions of $F/m_{\rm pl}$ and
$\alpha$, respectively. Similarly, it can be seen that hilltop
inflation in the large parameter region of $\mu$ also coincides with
the chaotic inflation model of $p=4/3$~\cite{Kohri:2014rja}.

\begin{figure}[!h]
  \centering
  \includegraphics[width=0.8\columnwidth]{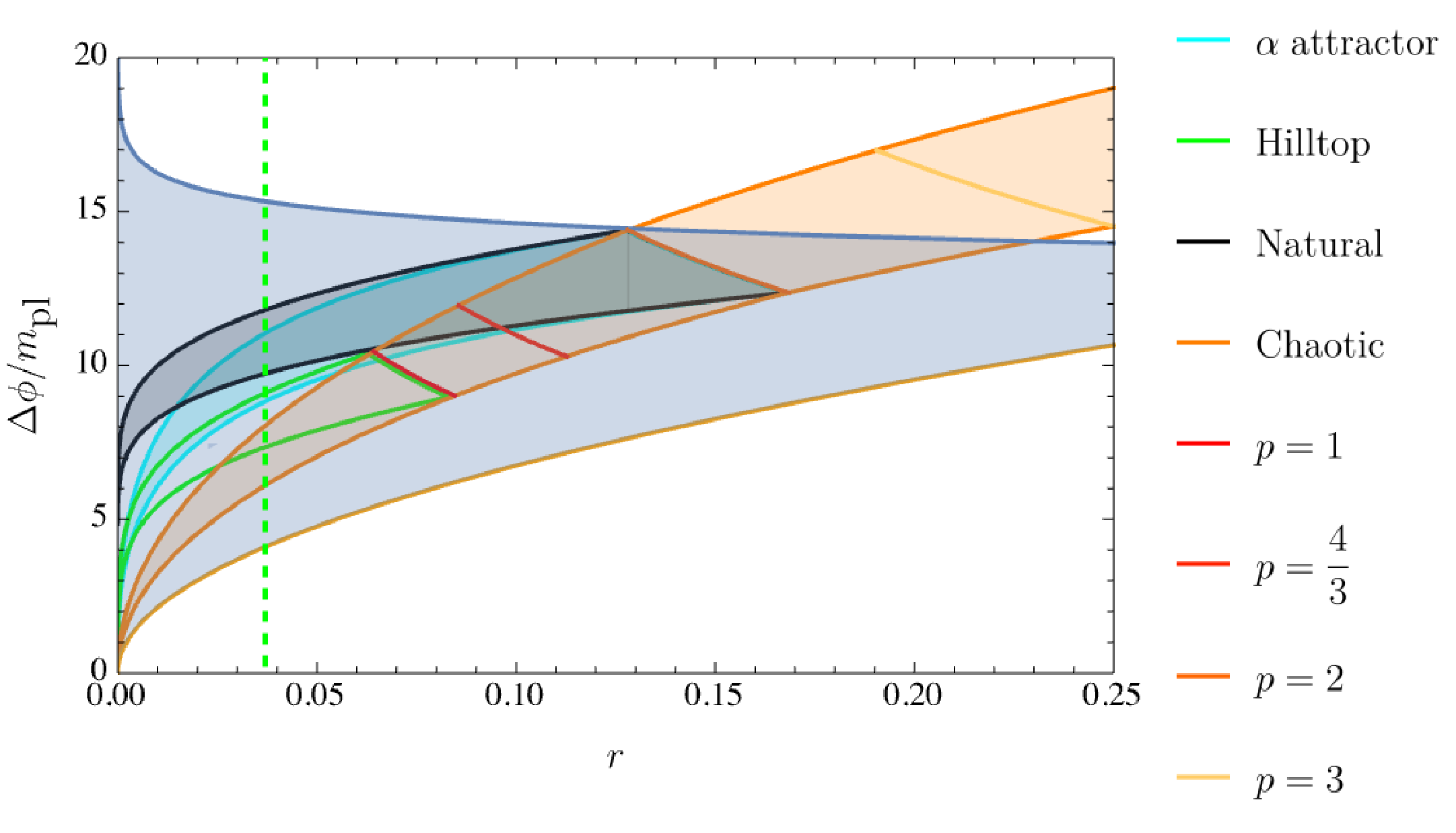}
  \caption{The same as that of Fig.~\ref{fig:sdc_lb_e1}, but for $\lambda_{\mathrm{dc}} =\sqrt{\frac12}$.}
  \label{fig:sdc_lb_u1}
\end{figure}

\begin{figure}[!h]
  \centering
  \includegraphics[width=0.8\columnwidth]{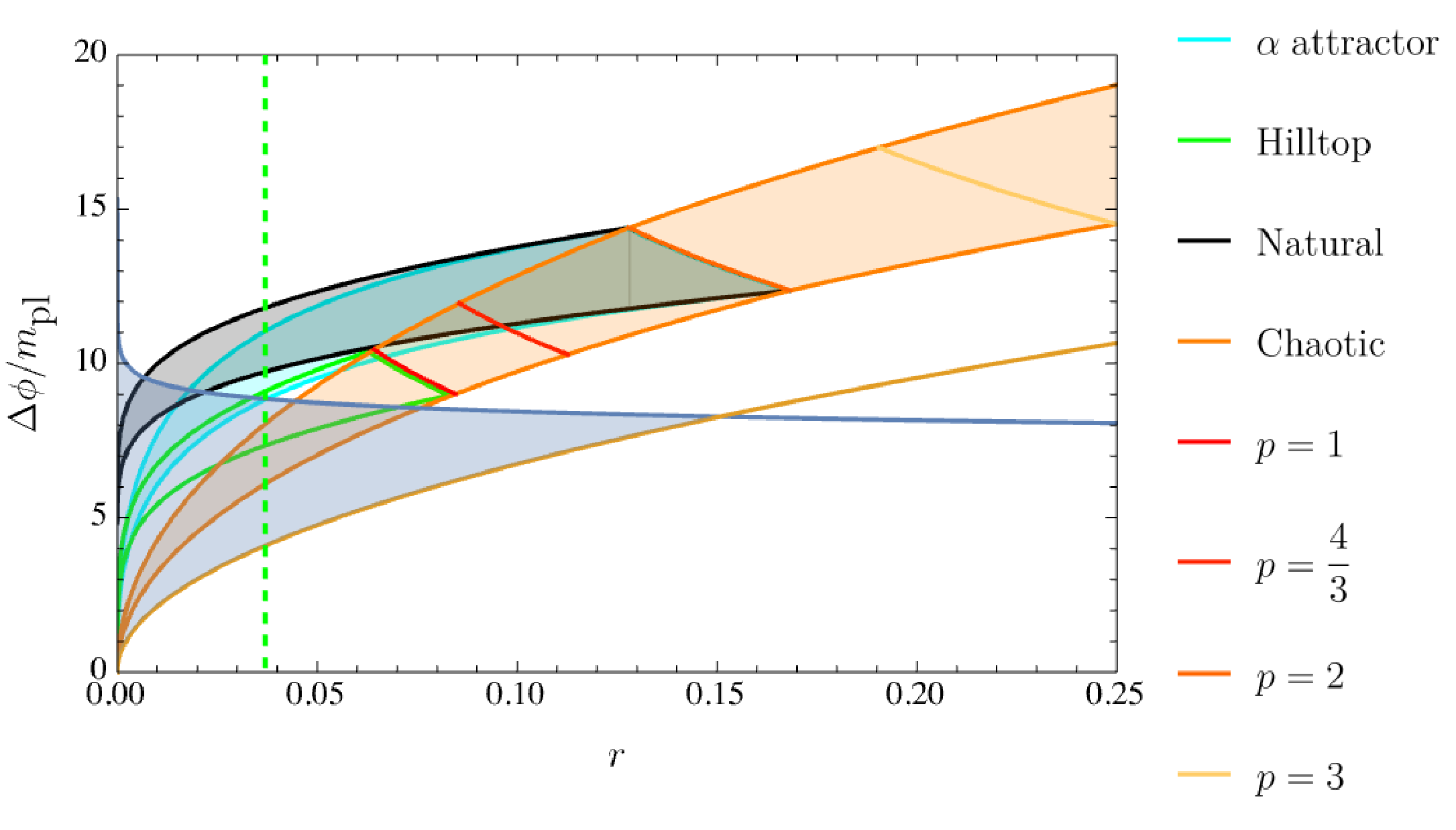}
  \caption{The same as that of Fig.~\ref{fig:sdc_lb_e1}, but for $\lambda_{\mathrm{dc}} =\sqrt{3/2}$.}
  \label{fig:sdc_lb_o1}
\end{figure}

In Fig. \ref{fig:sdc_lb_u1}, we chose
$\lambda_{\mathrm{dc}} = 1/\sqrt{2}$, which is the case for the tower
of moduli space. On the other hand, in Fig.~\ref{fig:sdc_lb_o1}, the
case of $\lambda_{\mathrm{dc}} =\sqrt{3/2}$, which corresponds to the
case of the KK tower.
It is interesting that the case for
$\lambda_{\mathrm{dc}} =\sqrt{3/2}$ gives the most stringent upper
bounds on $r$ and $\Delta \phi$. In this case, it is notable that only
the swampland distance conjecture  excluded some regions of $r$ at
around $r \sim 0.030$-$0.036$ and
$\Delta \phi \sim 9$-$10 m_{\rm pl}$, which is stronger than the
observational bound from the CMB by the Planck collaboration 2018,
$r < 0.036$ (the vertical green dashed
line)~\cite{BICEP:2021xfz,Tristram:2021tvh}. Only by such a
theoretical requirement, the models of the inflation such as the
natural inflation, the hilltop inflation and the $\alpha$ attractors
were killed by the swampland distance conjecture 
(Fig.~\ref{fig:sdc_lb_o1}).

\begin{figure}[!h]
  \centering
  \includegraphics[width=0.8\columnwidth]{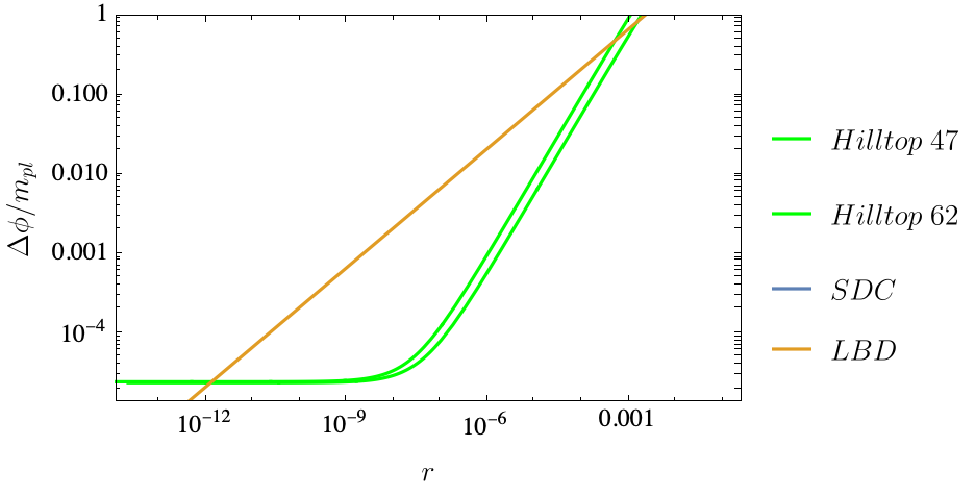}
  \caption{Enlarged figure for the hilltop inflation models (green
    solid lines) in the ($r$, $\Delta\phi$) plane. Two curves are
    plotted for $N$=47 (upper one) and $N$=62 (lower one),
    respectively. The orange line represents the Lyth bound.}
  \label{fig:vio}
\end{figure}

Finally, in Fig.~\ref{fig:vio}, we plot the enlarged figure of the
hilltop inflation models for smaller $\Delta \phi \ll m_{\rm pl}$ and
smaller $r \ll 10^{-3}$. Two curves are plotted for $N$=47 (lower one)
and $N$=62 (upper one), respectively. As shown in this figure, the
Lyth bound (orange line) does not work for the hilltop inflation
models in the small $r$ region, ranging from $10^{-12}$ to
$10^{-3}$, corresponding to $\mu_4$ between 0.024 and 4.93. The Lyth bound would be a good rough guide for observing
the general behavior of the prediction of $r$ in inflation
models. However, it sometimes gives a wrong prediction.  The results
of this study clearly show that, when comparing with observed values,
it is necessary to calculate the concrete predictions for each model,
not just by using the Lyth bound.

\section{Conclusions}
\label{sec:conc}

In this paper, we have studied the swampland conjecture, which has been
obtained in the construction of quantum gravity theory, to address the
question of how much we can narrow down the large field inflation
models to those that are quantum gravity-theoretically advantageous.
Among the swampland conjectures, the property known as the swampland
distance conjecture provides an upper bound on the primary
gravitational waves produced by large field inflation models.  On the
other hand, the Lyth bound, which provides a lower bound on primary
gravitational waves, plays a complementary role to the swampland
distance conjecture.  In this paper, we report the following new
points for the first time:

In terms of the upper bound on gravitational waves, some of the major
large field inflation models are more strongly constrained by the
swampland distance conjecture than that of the CMB constraints. It is
interesting that the case for $\lambda_{\mathrm{dc}} =\sqrt{3/2}$
gives the most stringent upper bounds on $r$ as a function of
$\Delta \phi$.\footnote{As remarked before, this is preferred from string theory point of view.} In this case, it is notable that the swampland distance
conjecture alone excluded $r$ at around $r \sim 0.030$-$0.036$ and
$\Delta \phi \sim 9$-$10 m_{\rm pl}$, which is stronger than the
observational bound from the CMB by the Planck collaboration 2018,
$r< 0.036$~\cite{BICEP:2021xfz,Tristram:2021tvh}. Only by such a
theoretical requirement, some parameters in the models of the
inflation, such as natural inflation, the hilltop inflation or the
$\alpha$ attractors are excluded by the swampland distance conjecture.

We also have found the model parameters that violate the Lyth bound
for certain parameter regions of the hilltop inflation model.
Actually, the Lyth bound cannot satisfy the parameters of the hilltop
inflation models in the small $r$ region, ranging from $10^{-12}$ to
$10^{-3}$, corresponding to $\mu_4$ between 0.024 and 4.93. When we compare the predictions by a model of the inflation
with the observed values, our results clearly show that we have to
calculate the concrete predictions for each model, not just by using
the Lyth bound.

The latest results from the Atacama Cosmology Telescope (ACT)\cite{ACT:2025fju, ACT:2025tim} show a slightly larger spectral index compared to Planck 2018. At the CMB pivot scale $k_{\rm{CMB}}$ =0.05$\rm{Mpc}^{-1}$, the spectral index was constrained to be $n_s = 0.974 \pm 0.003$ (68\% CL). This result excludes the natural inflation model at more than the 2$\sigma$ level. Furthermore, although the $\alpha$-attractor model showed good agreement with Planck 2018 regarding the tensor-to-scalar ratio and spectral index, the ACT partially allowed for the $e$-foldings number $N \gtrsim 57$. Many other parameter regions are under tension at the 2$\sigma$ level.
 
In contrast, even though the hilltop inflation model is excluded for small parameter $\mu_4$ regions, i.e., $\mu_4/m_{pl} \lesssim 14$, the chaotic inflation model and the hilltop inflation model are allowed over large regions according to these new results, compared to the previous two models. In particular, the shift towards a larger spectral index by the latest ACT has opened up parameter regions where the parameter $p$ of the chaotic inflation model is less than 1. Interestingly, our results are able to constrain the region where $p<1$, as shown in Fig. 5. This means that the ACT and our results are complementary to each other.

\section*{Acknowledgments}

This work was in part supported by JSPS KAKENHI Grants No.~JP24H00976 (Y.H.), No.JP24K07035 (Y.H.), No.JP24KF0167 (Y.H.) No.23KF0289 (K.K.), and No.~JP24K07027 (K.K.), and MEXT KAKENHI Grant No.~JP24H01825 (K.K.).

\section*{DATA AVAILABILITY}
No data were created or analyzed in this study.

\appendix



\bibliography{apssamp}

\end{document}